\documentclass[twocolumn,journal]{IEEEtran}
\usepackage[T1]{fontenc}
\usepackage[latin9]{inputenc}
\usepackage{array}
\usepackage{multirow}
\usepackage{amsmath}
\usepackage{amssymb}
\usepackage{graphicx}
\usepackage[unicode=true,
 bookmarks=true,bookmarksnumbered=true,bookmarksopen=true,bookmarksopenlevel=1,
 breaklinks=false,pdfborder={0 0 0},pdfborderstyle={},backref=false,colorlinks=false]
 {hyperref}
\hypersetup{pdftitle={Your Title},
 pdfauthor={Your Name},
 pdfpagelayout=OneColumn, pdfnewwindow=true, pdfstartview=XYZ, plainpages=false}

\makeatletter

\providecommand{\tabularnewline}{\\}

\let\oldforeign@language\foreign@language
\DeclareRobustCommand{\foreign@language}[1]{%
  \lowercase{\oldforeign@language{#1}}}

\usepackage[caption=false,font=footnotesize]{subfig}
\usepackage{lipsum}
\usepackage{mathtools}
\usepackage{cuted}
\usepackage{dsfont}

\makeatother

\begin{document}
\title{Contact Classification in\\
COVID-19 Tracing}
\author{Christoph Günther \thanks{Christoph Günther is with the German Aerospace Center, 82234 Weßling,
and with Technische Universität München, 80330 Munich, Germany, e-mail:
\protect\href{http://KN-COVID\%40dlr.de}{KN-COVID@dlr.de}.}and Daniel Günther\thanks{Daniel Günther is a student at Technische Universität München, 80330
Munich, Germany, e-mail: \protect\href{http://d.guenther\%40tum.de}{d.guenther@tum.de}.} }
\markboth{}{}
\IEEEpubid{}
\maketitle
\begin{abstract}
The present paper addresses the task of reliably identifying critical
contacts by using COVID-19 tracing apps. A reliable classification
is crucial to ensure a high level of protection, and at the same time
to prevent many people from being sent to quarantine by the app. Tracing
apps are based on the capabilities of current smartphones to enable
a broadest possible availability. Existing capabilities of smartphones
include the exchange of Bluetooth Low Energy (BLE) signals and of
audio signals, as well as the use of gyroscopes and magnetic sensors.
The Bluetooth power measurements, which are often used today, may
be complemented by audio ranging and attitude estimation in the future.
Smartphones are worn in different ways, often in pockets and bags,
which makes the propagation of signals and thus the classification
rather unpredictable. Relying on the cooperation of users to wear
their phones hanging from their neck would change the situation considerably.
In this case the performance, achievable with BLE and audio measurements,
becomes predictable. Our analysis identifies parameters that result
in accurate warnings, at least within the scope of validity of the
models. A significant reduction of the spreading of the disease can
then be achieved by the apps, without causing many people to unduly
go to quarantine. The present paper is the first of three papers which
analyze the situation in some detail.
\end{abstract}


\IEEEpeerreviewmaketitle{}

\section{Introduction}

\IEEEPARstart{T}{he} COVID-19 pandemic has spread to enormous dimensions
with 16 Million people affected and more than 644'000 fatalities up
to July 26th, 2020. Unfortunately, the rate of increase has only flattened
in China and selected European countries. The most important effective
method to slow down the pandemic has so far been the enforcement of
quarantine to large portions of the population, which led to a massive
economic disruption. In countries such as China, South Korea, Singapore,
and a number of European countries, the reduced infection rates made
it possible to alleviate some of the restrictions. This involves the
obligation to use masks and at least a recommendation to use some
form of contact tracing. Different proposals for such a tracing have
been made \cite{Kupf20} and several different approaches are being
followed in various countries. The most interesting proposals are
those that fully focus on the tracing of contacts without tracking
the movement of individuals, such as the scheme implemented in Germany
\cite{Coro20}. The associated concepts were developed nearly synchronously
by a number of authors and were published in \cite{ArBecBlanCol20},
\cite{GueGueGue20}, and \cite{NanAndBarBol20}. A review of associated
requirements is found in \cite{RasSchBarVilc20} and a review of major
apps in \cite{ONeRyaJoh20}. In view of the highly contagious nature
of COVID-19, of the lack of a vaccine and of the high casualty rates,
an effective tracing and significant testing capabilities are essential.
In Germany 16 million people have downloaded the associated app on
their iPhone and Android Phones so far.

Tracing apps rely on Bluetooth to detect the proximity of other people's
devices. These apps generate random IDs, which are broadcast and stored
to identify contacts in the case that the owner of a device is tested
positively. If the owner is tested positively, the list of IDs stored
on his device is published. Conversely, each device keeps the IDs
of past contacts and compares them to the published list of IDs on
a regular basis in order to establish whether a critical contact has
taken place. Apple published an update of its operating system to
support the development of such apps (iOS 13.1.5) and Google updated
its Application Programming Interface (API). In the case of a critical
contact the person should quarantine himself and register for testing.
The outcome might be that he is found to be a carrier of the disease.
In this case, the owner should trigger the release of his device's
list of random IDs. The consequences of positive and negative testing
depends on local regulations. In Germany a contact is characterized
as critical and is called a Category 1 contact, if two people were
in a face-to-face meeting at a distance of less than 2 meters for
more than 15 minutes. The present paper relies on this definition
but its parameters can easily be adapted to any other definition.

Several countries have released tracing apps. The classification methods
used are typically not discussed publicly. Ideally, the classification
ensures a minimum missed detection rate at an acceptable level of
false alarms. In the case of too many false alarms, people will be
unduly sent to quarantine and the app-based will be rejected by the
public. If on the other side the app fails to identify potential carriers,
they continue infecting others and its effectiveness is jeopardized.
As shall be seen both issues are most critical in the case of a high
densities of COVID-19 carriers. As a consequence, the present analysis
will be most pertinent to regions with a high infection rate. This
paper is the first of a series of three papers. The other two papers
address the particularities of the evaluation of Bluetooth Radio Signal
Strength Indicators (BT-RSSI) \cite{DamGueGenGue20} and of audio
ranging \cite{KurGueGue20} in more detail.

Electromagnetic signals, such as Bluetooth signals, can be used for
time of flight measurements, which provides accurate ranging results.
Unfortunately, this and some other ideas cannot be considered presently,
since a contact tracing app must rely on existing smartphones and
devices. Thus, only existing functions provided by the chipsets, and
even more importantly by the APIs of the devices can be used. The
options for Bluetooth on existing equipment are limited to power measurements.
The outcome of such measurements very much depend on the location
of the device, which might be in a pocket or in a bag, often together
with keys, coins, metallic business card holders and the like. Furthermore,
the human body, with a strong water content, strongly absorbs Bluetooth
signals. Together these uncertainties greatly influence the power
levels measured at a distant receiver. The difficulties of tracing
contacts by Bluetooth power measurements are also discussed in \cite{ONei20}.
The remaining uncertainties about a potential contact to an infected
person could potentially be resolved by interrogating the people involved.
This would require the disclosure of the location at the time of contact,
which might have been on a commuter train or at lunch in a restaurant,
for example. The people must then identify where they sat or stood,
which they might remember or not. In any case, this would be a source
of privacy issues, discomfort and residual uncertainty. The German
app would not support such a manual tracing anyway, since it does
not collect the necessary information! In any case, such a human intervention
would reduce the level of acceptance.

\begin{figure}[tbh]
\begin{centering}
\includegraphics[scale=0.2]{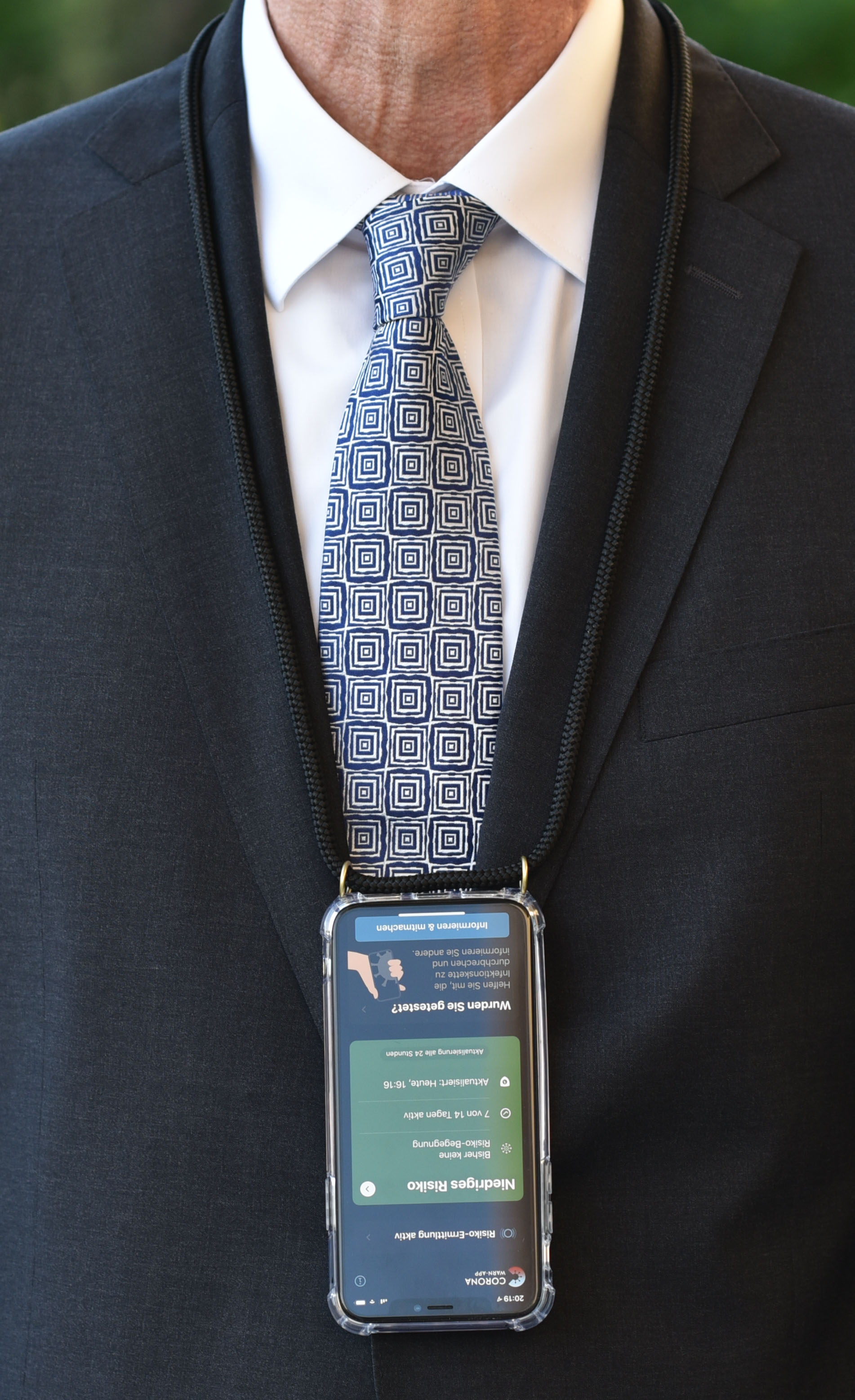}
\par\end{centering}
\caption{Example of a hull for carrying a smartphone hanging from the neck.\label{fig:Handykette}}

\end{figure}

As a consequence, we propose to carry the smartphone in an exposed
manner, namely hanging around one's neck. In summer time, younger
people often do that already. On the basis of the present findings,
this is recommended to everyone, also in a business context, see Figure
\ref{fig:Handykette}. Corresponding cases are available from several
vendors. This mode of wearing the smartphone ensures a line of sight
situation between two fellows facing each other. It leads to measurements
that are a lot easier to interpret using Bluetooth Radio Signal Strength
Indication (RSSI), audio ranging as well as gyro and magnetic sensors.
The paper starts with a description of the statistical relationship
between individual measurements and their classification in Section
\ref{sec:Statistical-Considerations}. This section lays the foundation
for evaluating the performance of classification in simulations or
experiments. The probability of missed detection turns out to be critical
for the success of the classification. Bluetooth RSSI evaluation is
rather sensitive to the manner in which measurements are evaluated.
Section \ref{sec:Power-Measurements} describes some aspects relating
to the modeling of Bluetooth propagation and power measurements, as
well as the essential result from the more in-depth study of the situation
developed by Dammann et al. \cite{DamGueGenGue20}. The following
Section \ref{sec:Acoustic-Ranging} addresses audio ranging, which
turns out to be an important complementary technique. Some audio properties
of smartphones are summarized in this section. A more detailed study
is published by Kurz et al. \cite{KurGueGue20}. Section \ref{sec:Attitude-Sensing}
shortly addresses the possibility of using attitude sensing, which
is not explored in depth. Section \ref{sec:Classification} finally
discusses some basics of classifying contacts using the set of sensors
mentioned.

\section{Statistics of Classification\label{sec:Statistical-Considerations}}

The success of classifying contacts into Category 1 and other contacts
depends critically on our capability of estimating distances. As a
consequence, it is important to understand the influence of under-
and overestimating distances from a pandemic point of view. This requires
a study of the associated statistics. For a Category 1 contact, two
fellows have to be facing each other at a distance of less than 2
meters for at least 15 minutes. This is called a $C_{1}$ contact
throughout the paper.

Assume that we are the person A and that we monitor the presence of
B. We aim at determining whether the contact to B is a $C_{1}$ contact
or not, denoted by $C_{1}$ or $\neg C_{1}$, respectively. Furthermore,
denote the outcome of the estimation process by $\hat{C}_{1}$ and
$\neg\hat{C}_{1},$ then there are four different possibilities, as
listed in Table~\ref{tab:C1-Contact-Events} (classical hypothesis
testing).

\begin{table*}[tbh]
\caption{C1 Contact Events\label{tab:C1-Contact-Events}}

\centering{}%
\begin{tabular}{c|c|c}
Event & Description & Probability\tabularnewline
\hline
$\hat{C}_{1}|C_{1}$ & $C_{1}$ is detected and $C_{1}$ is correct ($C_{1}$ contact) & $p_{d}$\tabularnewline
$\neg\hat{C}_{1}|C_{1}$ & $C_{1}$ is rejected while $C_{1}$ would have been correct (missed
detection) & $p_{md}=1-p_{d}$\tabularnewline
$\hat{C}_{1}|\neg C_{1}$ & $C_{1}$ is detected but $C_{1}$ is not correct (false alarm) & $p_{fa}$\tabularnewline
$\neg\hat{C}_{1}|\neg C_{1}$ & $C_{1}$ is rejected, which is the correct conclusion(no contact) & $1-p_{fa}$\tabularnewline
\end{tabular}
\end{table*}

Obviously, in any good design $p_{md}$, and $p_{fa}$ are small.
The four cases have to be considered jointly with the possibility
that B is tested positively, which happens with probability $p_{i},$
and shall be denoted by $B.$ Current values, based on data published
by John Hopkins University on July 26th, for$p_{i}$ are $1/5'100$
for Germany and $1/113$ for the USA. If B is either not tested
or tested negatively, this shall be denoted by $\neg B$. Let finally
$p_{C_{1}}$ be the probability of $C_{1},$ then this leads to the
situations summarized in Table~\ref{tab:C1-Contact-Event}.

\begin{table*}[tbh]
\caption{$C_{1}$ Contact Event Probabilities and Consequences\label{tab:C1-Contact-Event}}

\centering{}%
\begin{tabular}{c|c|c}
Event & Probability & Consequence\tabularnewline
\hline
$\hat{C}_{1}|C_{1}\land B$ & $p_{C_{1}}p_{i}p_{d}\sim p_{C_{1}}p_{i}$. & A goes to quarantine\tabularnewline
$\neg\hat{C}_{1}|C_{1}\land B$ & $p_{C_{1}}p_{i}p_{md}$ & A pot. spreads the virus\tabularnewline
$\hat{C}_{1}|\neg C_{1}\land B$ & $(1-p_{C_{1}})p_{i}p_{fa}$ & Unnecessary quarantine of A\tabularnewline
$\neg B\lor(\neg\hat{C}_{1}|\neg C_{1}\land B)$ & $1-p_{i}+(1-p_{C_{1}})p_{i}(1-p_{fa})$ & none (all other cases)\tabularnewline
\end{tabular}
\end{table*}

The first and fourth rows of Table~\ref{tab:C1-Contact-Event} provide
the desired outcome. The probability $p_{C_{1}}$ of a contact being
$C_{1}$ is driven by social behavior. Social distancing aims at reducing
$p_{C_{1}}.$ This is important, since many people would otherwise
be potentially infected and sent to quarantine by the first row in
the table, whenever $p_{i}$ is significant. The product $p_{C_{1}}p_{i}$
is the probability that the contact is $C_{1}$ and that fellow B
is infected at the same time. Aiming at a small value of $p_{md}$
ensures that few potential carriers continue spreading the disease
(second row). The actual value of $p_{md}$ is a direct measure of
the containment benefit provided by a tracing app. Since $1-p_{C_{1}}$
is large, it is very important that the probability $p_{fa}$ of wrongly
classifying a contact as being $C_{1}$ be small. Otherwise, numerous
people would be unduly sent to quarantine by the third row. The value
of $p_{fa}$ characterizes the extra load in terms of quarantining
and testing generated by a tracing app. This has to be taken into
account in the trade-off of $p_{fa}$ versus $p_{md}$. Also note
that the undesired outcomes, i.e. the rows 2 and 3, have a probability
proportional to $p_{i}$, which means that they are unlikely to occur
in the case of a small density of infectious people. As a consequence,
a potential under-performance of an app only becomes apparent in environments
with a high number of infectious people.

The decision for $\hat{C}_{1}$ or $\neg\hat{C}_{1}$ is taken after
a substantial number of individual measurements. They are assumed
to be performed at regular intervals. The number of such intervals
in a time laps of 15 minutes is denoted by $x_{0}.$ Depending on
the assumed behavior of people, different methods of analyzing the
measurement data shall be considered:
\begin{itemize}
\item Model A: People are rather mobile and the environment is changing
quickly - the contact duration is accumulated over many short intervals.
Examples of such situations occur when people work closely together,
which is not particularly critical in terms of classification. They
occur in underground trains, during breaks at conferences, at any
form of party and the like. In these cases, a decision is taken every
15 seconds, if $x_{0}=60$ such measurements indicate that fellow
B is in the contact zone of fellow A the contact is classified as
being $C_{1}$. It will turn out that this model cannot be addressed
with the current capabilities.
\item Model B: People come together, stay in a given relative pose and then
separate again. This happens when people are seated in a train, especially
in long-distance inter-city trains, in restaurants, meeting rooms,
lecture halls, theaters and the like. In this case, a single test
$(x_{0}=1)$ is performed to decide on whether A is in the contact
zone of B. Specifically, in the case of Bluetooth RSSI measurements,
a timer is started when the RSSI value exceeds a critical value for
the first time. From then on, the times for which the RSSI values
are compatible with a $C_{1}$ contact are accumulated. If the time
exceeds 15 minutes at the end of the contact, a $C_{1}$ contact is
declared. There are many different options for the implementation
of this model. They will not be further discussed, however, since
they assumes a static constellation of people, which is not the most
common case.
\item Model C: Is an intermediate model, which allows for slow changes in
the distribution of people. In this model the RSSI is accumulated
over time, like in Model B but only over intervals of 3 or 5 minutes.
Contrary to Model B, there is no further condition on the accumulation.
The accumulated RSSI-values are evaluated against a threshold at the
end of the interval. Exceeding the threshold $x_{0}=5$ times (3 minutes
of accumulation) or 3 times (5 minutes of accumulation), leads to
the decision $\hat{C}_{1}$. This approach is more robust with respect
to the behavior of people and preferable to Model B.
\end{itemize}
Model A is most universally valid with respect to people's behavior.
Its statistics are so unfavorable that it does not lead to acceptable
values of $p_{md}$, however. In all models, the number of RSSI measurement
$n$ that are combined, before taking an elementary decision, is another
parameter that can be adapted. Large values lead to more reliable
decisions but also to a higher number of exchanged messages. The rate
of message will be $n\cdot x_{0}$ measurements in 15 minutes.

In order to assess $p_{md},$ we need to know the number of times
that the distance and attitude condition for $C_{1}$ between A and
B are fulfilled. This depends on the profession and personality of
the person. It has two components, the first one is determined by
the number of people met during one day. Let us assume that this number
is $k$ and that it has probability $p_{n}(k)$, then the probability
$p_{S}$ that a particular fellow A spreads the virus after having
been in contact with $m\in\left\{ 1,2,\ldots k\right\} $ people,
who are infectious with probability $p_{i},$ under the assumption
that $i\in\left\{ 1,2,\ldots m\right\} $ of these contacts are not
detected, is given by:
\begin{align*}
p_{S} & =\sum_{k=0}^{\infty}p_{n}(k)\sum_{m=1}^{k}\left(\begin{array}{c}
k\\
m
\end{array}\right)p_{i}^{m}(1-p_{i})^{k-m}\\
 & \quad\cdot\sum_{i=1}^{m}\left(\begin{array}{c}
m\\
i
\end{array}\right)p_{md}^{i}(1-p_{md})^{m-i}.
\end{align*}
Since $p_{i}$ and $p_{md}$ are small numbers, the dominant term
in this equation is obtained for $m=i=1:$
\begin{equation}
p_{S}\simeq Kp_{i}p_{md},\label{eq:p_Disease_spreading0}
\end{equation}
with $K=\sum_{k=0}^{\infty}p_{n}(k)k$ being the average number of
contacts, see also the second row in Table \ref{tab:C1-Contact-Event}.
All these contacts take place mostly independently and can thus be
treated as such. Each of them is associated with a contact time $x\in\mathbb{N}$,
with a distribution $p_{X}(x)$. The latter is derived from social
models and depends on whether people are practicing social distancing.

The accumulation of $n$ measurements leads to a decision $\hat{c}_{1}$.
The latter has a probability of missed detection and false alarm denoted
by $\pi_{md}$ and $\pi_{fa}$, respectively. In the present section,
both quantities are written without further indices. In later sections,
the dependency on $n$ will be made explicit. The combination of $x_{0}$
such decisions $\hat{c}_{1}$ finally leads to the decision $\hat{C}_{1}$,
which is associated with a missed detection probability:

\begin{equation}
p_{md}(x)=\sum_{m=0}^{x_{0}-1}\left(\begin{array}{c}
x\\
m
\end{array}\right)(1-\pi_{md})^{m}\pi_{md}^{x-m},\label{eq:p_d}
\end{equation}
since the combined missed detection occurs whenever less than $x_{0}$
detections succeed. Using this in Equation \ref{eq:p_Disease_spreading0}
implies that the probability that A spreads the disease is:
\begin{equation}
\begin{split}p_{S} & \simeq Kp_{i}\sum_{x=x_{0}}^{x_{M}}p_{X}(x)p_{md}(x)\\
 & =Kp_{i}\sum_{x=x_{0}}^{x_{M}}p_{X}(x)\sum_{m=0}^{x_{0}-1}\left(\begin{array}{c}
x\\
m
\end{array}\right)(1-\pi_{md})^{m}\pi_{md}^{x-m},
\end{split}
\label{eq:p_Disease_Carrier}
\end{equation}
with, $x_{M}=24\cdot4\cdot x_{0}$ being the number of elementary
decisions taken per day ($24\cdot4$ quarters of hours times $x_{0}$).
The above equation is an approximation since the distribution of contact
times depends on the people and circumstances of the meeting, like
sitting together in the train, having a joint lunch and so on. If
$\pi_{md}\ll1,$ the term $m=x_{0}-1$ is dominant in Equation \eqref{eq:p_Disease_Carrier}:
\begin{align}
p_{S} & \sim Kp_{i}\sum_{x=x_{0}}^{x_{M}}p_{X}(x)\left(\begin{array}{c}
x\\
x_{0}-1
\end{array}\right)\pi_{md}{}^{x_{0}-1}\pi_{md}^{x-x_{0}+1}\nonumber \\
 & =Kp_{i}\pi_{d}^{x_{0}-1}\sum_{x'=1}^{x_{M}-x_{0}+1}p_{X}(x'+x_{0}-1)\label{eq:p_Disease_Carrier2}\\
 & \qquad\qquad\qquad\qquad\cdot\left(\begin{array}{c}
x'+x_{0}-1\\
x_{0}-1
\end{array}\right)\pi_{md}^{x'}\nonumber \\
 & \leq Kp_{i}\pi_{d}^{x_{0}-1}\sum_{x'=1}^{x_{M}-x_{0}+1}p_{X}(x'+x_{0}-1)\frac{1}{x'!}\left(x_{0}\pi_{md}\right)^{x'},\nonumber
\end{align}
with $\pi_{d}=1-\pi_{md}.$ The second line in the equation is obtained
by shifting the indices, the third one is obtained by expanding the
binomial coefficients and bounding the terms in the numerator. Note
that the term for $x'=1$ holds with equality. Under the same assumptions
used so far, the probability that fellow A is a $C_{1}$ contact of
B after a day is:
\[
p_{C_{1}}=Kp_{i}\sum_{x=x_{0}}^{x_{M}}p_{X}(x)=Np_{i}\sum_{x'=1}^{x_{M}-x_{0}+1}p_{X}(x'+x_{0}-1).
\]
Thus, the comparison of $p_{S}$, i.e. the probability of spreading
the virus with tracing, and of $p_{C_{1}}$, i.e. the corresponding
probability without tracing, shows that contact tracing is a very
effective option to reduce the spreading whenever
\begin{equation}
x_{0}\pi_{md}\label{eq:condition_pmd}
\end{equation}
is small. This implies that the probability of missed detection must
be constrained to a value smaller than $1/x_{0},$ which is possible
to achieve if $x_{0}$ is small, as it is the case in Model B and
Model C and not possible to achieve in Model A, even with very large
values of $n$. Rephrasing this in words may help developing some
intuition: since $x_{0}$ individual $\hat{c}_{1}$ decision are needed
for a $\hat{C}_{1}$ decision, missing any one of them leads to a
missed detection. Since there are $x_{0}$ options for that, $p_{S}$
becomes essentially proportional to $x_{0}\pi_{md}.$ We will use
the latter product as a measure for the reduction in the spreading
of the disease by the tracing app.

In order to evaluate $p_{fa}$, we need to additionally know the number
of times $y$ that a person is close enough for a measurement to take
place. The distribution $p_{Y}(y)$ does again depend on social parameters
but additionally depends on radio propagation in the case of Bluetooth
measurements, and on the triggering mechanism in the case of audio
ranging. The number of contacts $K_{Y}\geq K$ is larger, since the
presence detection by Bluetooth signaling is triggered well beyond
$C_{1}$ separation. Consider Bluetooth measurements: if among the
$y$ time instances for which a radio contact to one particular fellow
B persists, and assume that $m<x_{0}$ of those contacts are correctly
detected as fulfilling the $C_{1}$ conditions. Then, $q$ additional
erroneously identified contacts (erroneous $\hat{c}_{1}$-decisions)
with $m+q\geq x_{0}$ are needed to cause a false alarm for that number
$y$ of radio contacts to B (see Table \ref{tab:Variables-p_fa} for
a summary of the meaning of the variables):
\begin{equation}
\begin{split}p_{fa}(y)= & \sum_{x=0}^{x_{M}}p_{X}(x)\sum_{m=0}^{\min\{x,x_{0}-1\}}\left(\begin{array}{c}
x\\
m
\end{array}\right)(1-\pi_{md})^{m}\pi_{md}^{x-m}\\
 & \quad\sum_{q=x_{0}-m}^{y-x}\left(\begin{array}{c}
y-x\\
q
\end{array}\right)\pi_{fa}^{q}(1-\pi_{fa})^{y-x-q},
\end{split}
\label{eq:p_fa}
\end{equation}
for $y\geq x_{0}$ and $p_{fa}(y)=0$ for $y<x_{0}.$
\begin{table}[tbh]
\caption{Variables used in Equation \ref{tab:Variables-p_fa}.\label{tab:Variables-p_fa}}

\centering{}%
\begin{tabular}{c|c}
Variable & Meaning\tabularnewline
\hline
$y$ & number of radio contacts\tabularnewline
$x$ & number of $C_{1}$ contacts\tabularnewline
$x_{0}$ & number of $\hat{c}_{1}$-decisions to declare $C_{1}$\tabularnewline
m & number of correct $\hat{c}_{1}$ estimates\tabularnewline
q & number of incorrect $\hat{c}_{1}$ estimates\tabularnewline
\end{tabular}
\end{table}

Using Equation \eqref{eq:p_fa}, the expected number of an unnecessary
quarantining of people is approximated by:
\begin{equation}
\begin{split}n_{Q} & =K_{Y}p_{i}\sum_{y=x_{0}}^{x_{M}}p_{Y}(y)p_{fa}(y)\\
 & =K_{Y}p_{i}\sum_{y=x_{0}}^{x_{M}}p_{Y}(y)\cdot\\
 & \quad\cdot\ \sum_{x=0}^{x_{M}}p_{X}(x)\sum_{m=0}^{\min\{x,x_{0}-1\}}\left(\begin{array}{c}
x\\
m
\end{array}\right)(1-\pi_{md})^{m}\pi_{md}^{x-m}\\
 & \quad\cdot\sum_{q=x_{0}-m}^{y-x}\left(\begin{array}{c}
y-x\\
q
\end{array}\right)\pi_{fa}^{q}(1-\pi_{fa})^{y-x-q}.
\end{split}
\label{eq:p_Quarantine}
\end{equation}
This equation also includes the possibility that users move with respect
to each other, which means that the conditions $C_{1}$ and $\neg C_{1}$
alternate as a function of time. If $C_{1}$ is fulfilled $\pi_{fa}=0,$
and if $\neg C_{1}$, the equation $\pi_{d}=0$ holds. At the border
of the $C_{1}$ domain, the two quantities change their role. This
implies that a small $p_{fa}$ near that border is associated with
a large $p_{md}\sim1-p_{fa}$ on the other side of the border. This
is uncritical if the distributions are very narrow - concentrated
around a value - as is the case for ranging, but becomes rather problematic
with Bluetooth signal power measurements, which show a very flat distribution.
Unless great precautions are taken the classification becomes unreliable.
Consider the case, that fellow B is outside of the $C_{1}$ zone of
fellow A, i.e. $p_{X}(0)=1$. Then $x=0$ for these measurements and
the equation becomes:
\begin{align*}
n_{Q} & =K_{Y}p_{i}\sum_{y=x_{0}}^{x_{M}}p_{Y}(y)\sum_{x=x_{0}}^{y}\left(\begin{array}{c}
y\\
x
\end{array}\right)\pi_{fa}^{x}(1-\pi_{fa})^{y-x}.
\end{align*}
Although terms with $x>x_{0}$ may be larger, the term $x=x_{0}$
gives us an idea of the scaling. Its asymptotic dependency can be
evaluated using Stirling's formula and $\lim_{y\to\infty}(y/(y-x_{0}))^{y}=e^{x_{0}}$:
\begin{align*}
\left(\begin{array}{c}
y\\
x_{0}
\end{array}\right)\pi_{fa}^{x_{0}} & \sim\sqrt{\frac{y}{2\pi x_{0}(y-x_{0})}}\left(\left(\frac{y}{x_{0}}-1\right)e\pi_{fa}\right)^{x_{0}}.
\end{align*}
This means that in the long term, it is the duration of the radio
contact $y$, which dominates the rate of quarantining people. Some
target figures for $\pi_{fa}$ can be obtained for a fully occupied
train, for example. In Germany's 2nd class setups, there are 4 seats
in one row on each side of a carriage, and around 10 rows in the carriage.
The range of Bluetooth reaches well beyond the next row forward and
backward. This means that $K_{Y}>24$ of which 4-8 are within the
contact zone and must thus be discounted, leading to an effective
value $K_{Y}=16$. The value $y$ itself is determined by the duration
of the common journey. For commuter trains we choose 15 and 30 minutes,
for inter-city journeys 1, 2, and 3 hours, which leads to $y/x_{0}=1,2,4,8,$
and $12.$ In such a train a carrier of the disease will send 4 people
to quarantine, thus it should be tolerable that 2 additional people
are sent to quarantine by false alarms as well. The value of $\pi_{fa}$
is then obtained by solving
\[
K_{Y}\left(\begin{array}{c}
y\\
x_{0}
\end{array}\right)\pi_{fa}^{x_{0}}=2.
\]
Numerical values of $\pi_{fa}$ are indicated in Table \ref{tab:Indication for pfa}.
They are the values that can be tolerated, leading to a 50\% increase
in the quarantining of people riding a German train. The situation
is rather uncritical on a short commuter train ride $\pi_{fa}<0.93$
and much more demanding on a longer intercity train journey.

\begin{table}[tbh]
\caption{Rough indication for acceptable values of $\pi_{fa}.$ They are obtained
by considering train rides.\label{tab:Indication for pfa}}

\begin{centering}
\begin{tabular}{|c|c|c|c|c|c|}
\hline
$y/x_{0}$ & 1 & 2 & 4 & 8 & 12\tabularnewline
\hline
$\pi_{fa}$ & 0.93 & 0.25 & 0.11 & 0.05 & 0.03\tabularnewline
\hline
\end{tabular}
\par\end{centering}
\end{table}

\section{Bluetooth Power Measurements\label{sec:Power-Measurements}}

The Application Programming Interfaces (API) of Android and iOS allow
to trigger the transmission of Bluetooth Low Energy (BLE) advertisement
messages and to measure the radio signal strength of the received
signals. The corresponding values are provided in the form of a Radio
Signal Strength Indicators (RSSI), which is defined as the received
signal power on a logarithmic scale. Bluetooth uses frequencies from
a band shared with microwave heating, which means that Bluetooth signals
are strongly absorbed by water. As a consequence any part of a human
body obstructing the line of sight significantly attenuates the signal.
The wide variety of options for carrying mobile phones in your hand,
pocket or bag thus implies an enormous variability in received power
levels. This is further amplified by the directional characteristic
of low-cost antennas. You might make an experiment yourself using
a Bluetooth module and a BLE scanner app on your smartphone, which
can be downloaded from the iOS or Android stores. With the module
and phone separated by 1.5 meters, I personally found the following
RSSI-values: -61 to -66 dBm when the module was in my hand and -81
to -89-dB when it was in my pocket. Knowing that a 20 dB change corresponds
to a factor 10 in distance exemplifies the difficulty of estimating
distances using Bluetooth RSSI values. This led us to propose the
rule of carrying smartphones hanging down from the neck. Note that
the smartphone could be replaced by a much smaller device built around
a Bluetooth module, an Inertial Navigation System (INS) and a sonic
or ultra-sonic ranging system, as well.

Even if people follow the above recommendation on how to carry their
smartphone, the situation remains difficult due to uncertainties in
radio propagation, which furthermore takes place on three different
carrier frequencies. The unknown association of carrier frequencies
to measurements adds an additional level of difficulty. Gentner et
al. identified certain patterns in the use of carriers, see \cite{GenGueKin20},
which can be used to reduce the associated uncertainty. Traditional
models of propagation are shortly addressed in the following section
and in more details in \cite{DamGueGenGue20}. The section furthermore
relates the associated statistics to the statistics of classification.

\subsection{Propagation Model\label{subsec:BLE-Propagation}}

The smartphone is assumed to be worn on the chest, see \cite{DamGueGenGue20}
for details of the measurement setup used to obtain numerical results.
For each individual carrier, the received signal power ${\cal P}_{RX}$
is modeled by the equation:
\begin{equation}
{\cal P}_{RX}=\frac{\gamma}{d^{\nu}}{\cal P}_{TX}+n,\label{eq:RX_Power}
\end{equation}
with ${\cal P}_{TX}$ denoting the transmit power, $\gamma$ denoting
a stochastic fading coefficient, $d$ being the distance between the
receiver and the transmitter, $\nu$ being the exponent of the decay
law, which is 2 for free space propagation, and with $n$ representing
a superposition of noise and interference. For simplicity, the noise
and interference are not further considered here - at low distances
they are not dominant. In this case, the received power, can be represented
on a logarithmic scale, which leads to the definition of the RSSI:
\begin{equation}
RSSI=10\log{\cal P}_{RX}=10\log{\cal P}_{TX}-\nu\cdot10\log d+\eta\label{eq:RSSI}
\end{equation}
with $\eta=10\log\gamma$ and with logarithms taken to the basis 10.
The relationship between the reported RSSI value and $d$ is the basis
for distance measurement: the measured RSSI is compared to

\[
\Theta=10\log{\cal P}_{TX}-\nu\cdot10\log d_{c}+\langle\eta\rangle,
\]
with $d_{c}=2$ m being the critical distance. Note that Equation
\eqref{eq:RX_Power} defines the units, which have to be maintained
after taking logarithms.

In order to evaluate the missed detection probability per event $p_{md}$
or the false alarm probability per event $p_{fa},$ the statistics
for $\eta$ or $\gamma$ need to be known. These statistics are dependent
on the situation. In the case, that two fellows face each other, they
are in a line of sight situation. If the direct path dominates all
other contributions, $\gamma$ is basically delta distributed with
an average of $\Gamma$ determined by the antenna pattern. In other
cases, the direct path remains present but is superposed by scattered
components. In this case, the distribution of the amplitude of the
received signal is modeled by a Ricean distribution. This model is
considered to provide a faithful representation of reality, whenever
the parameters are properly estimated. Presently the model is only
considered for comparative purposes, as shall be seen below. The received
power (or attenuation $\gamma)$ in this model has a non-central $\chi^{2}$-distribution
with two degrees of freedom:

\begin{equation}
p_{R}(\gamma)=\frac{1}{2\sigma_{R}^{2}}e^{-(\gamma+\gamma_{R})/(2\sigma_{R}^{2})}I_{0}\left(\frac{\sqrt{\gamma\gamma_{R}}}{\sigma_{R}^{2}}\right),\label{eq:p_R}
\end{equation}
with $\gamma_{R}$ being the non-centrality parameter and $\sigma_{R}$
being the variance. In the case that the decision about $C_{1}$ is
taken on the basis of a single measurement $(n=1)$, e.g. in Model
A, the criterion for the decision is:
\begin{equation}
\gamma\geq\gamma_{c}\left(\frac{d}{d_{c}}\right){}^{\nu},\label{eq:gamma_condition}
\end{equation}
with $\gamma_{c}$ being given by:
\begin{equation}
\gamma_{c}=\langle\gamma\rangle=\int_{0}^{\infty}d\gamma\gamma p_{R}(\gamma).\label{eq:gamma_c}
\end{equation}
The associated estimate is denoted by $\hat{c}_{1},$ and the probability
of missed detection for the distance $d<d_{c}$ is given by:
\begin{equation}
\pi_{md}(d)=\int_{0}^{\gamma_{c}\left(d/d_{c}\right)^{\nu}}d\gamma p_{R}(\gamma).\label{eq:pi_md}
\end{equation}
If one would add several power measurements, i.e. $n>1,$ e.g. in
Model B and C, this would mean adding $n$ independent identically
distributed variables, each of them being $\chi^{2}$-distributed
with 2 degrees of freedom. The result would then be $\chi^{2}$-distributed
with $2n$ degrees of freedom:
\[
p_{R,n}(\gamma)=\frac{1}{2\sigma_{R}^{2}}\left(\frac{\gamma}{n\gamma_{R}}\right)^{\frac{n-1}{2}}e^{-(\gamma+n\gamma_{R})/(2\sigma_{R}^{2})}I_{n-1}\left(\frac{\sqrt{n\gamma\gamma_{R}}}{\sigma_{R}^{2}}\right).
\]
The Equations \eqref{eq:gamma_condition} and \eqref{eq:gamma_c}
would remain valid and the latter integral could be computed in closed
form for arbitrary $n$. The value $\gamma_{c}$ is the first moment
of the $\chi^{2}$-distribution with 2n degrees of freedom and non-centrality
parameter $n\gamma_{R}/\sigma_{R}^{2}$:
\[
\gamma_{c}=n(\gamma_{R}+2\sigma_{R}^{2}).
\]
The probability of missed detection \eqref{eq:pi_md} in estimating
$\hat{c}_{1}$ could then be computed in closed form using Marcum's
Q-function $Q_{n}(.,.)$:

\begin{align}
\pi_{md,R,n}(d) & =1-Q_{n}\left(\frac{\sqrt{n\gamma_{R}}}{\sigma_{R}},\frac{\sqrt{\gamma_{c}}\left(\frac{d}{d_{c}}\right)^{\nu/2}}{\sigma_{R}}\right).\label{eq:pmd_d}
\end{align}

The above distributions are adequate for users A and B in close proximity
of each other, as is the case for $d\leq d_{c}.$ It is the desired
result in Model A and shall serve as a benchmark in the Models B and
C. The reason for not using this result directly in the latter models
is that apps are expected to add the RSSI values rather than the power
values. In this case, the statistics cannot be determined in closed
form but must rather be evaluated numerically. Before addressing this
case, let us consider the situaiton $d>d_{c}$ with a line of sight
that is often obstructed. In such cases, a lognormal fading distribution
is considered to be a reasonable model of reality, see \cite{Hash93}.
The distribution may either be written in terms of $\gamma$:

\[
p_{L}(\gamma)=\frac{10\log_{10}(e)}{\sqrt{2\pi}\sigma_{L}\gamma}e^{-(10\log\,\gamma-10\log\,\gamma_{L})^{2}/(2\sigma_{L}^{2})},
\]
or in terms of $\eta=10\log\gamma$:
\begin{equation}
p_{L}(\eta)=\frac{1}{\sqrt{2\pi}\sigma_{L}}e^{-(\eta-\eta_{L})^{2}/(2\sigma_{L}^{2})},\label{eq:Gaussian}
\end{equation}
with $\eta_{L}=10\log\gamma_{L}=\langle\eta\rangle.$ Equation \eqref{eq:Gaussian}
makes the Gaussian character and the meaning of $\eta_{L}$ and $\sigma_{L}$
obvious. In the above discussion, a decision in the case $n=1$ was
taken in favor of $C_{1}$, whenever the power level was above a threshold.
On the logarithmic scale this condition reads $RSSI>\Theta$, i.e.
whenever the difference
\begin{equation}
RSSI-\Theta=\eta-\eta_{L}+\nu\cdot10\log\frac{d_{c}}{d}\label{eq:RSSI-Theta}
\end{equation}
is positive or equivalently whenever $\eta>\langle\eta\rangle+\nu\cdot10\log(d/d_{c})$.
Thus, a false alert occurs if this condition is fulfilled for $d>d_{c}$.
The probability of a false alarm, i.e. and erroneous decision for
$c_{1}$, becomes
\begin{align}
\pi_{fa}(d) & =\int_{\langle\eta\rangle+\nu\cdot10\log(d/d_{c})}^{\infty}d\eta\,p_{L}(\eta)\label{eq:pfa_d}\\
 & =Q\left(\frac{\nu\cdot10\log(d/d_{c})}{\sigma_{L}}\right),\nonumber
\end{align}
with the present $Q$-function being a scaled version of the error
function complement:
\[
Q(x)=\frac{1}{2}\text{erfc}\left(\frac{x}{\sqrt{2}}\right).
\]
In the case of $n=1$, a closed form of the statistics thus exists
for $\pi_{md}$ for $d\leq d_{c}$ and for $\pi_{fa}$ for $d>d_{c}.$
In the case $n>1,$ e.g. Model B and C, the situation changes somewhat
since measurements are now combined by adding RSSI-values. This corresponds
to a geometric average of the received powers. In this case, the probability
of false alarm can be computed easily:
\begin{equation}
\pi_{fa,n}(d)=Q\left(\frac{\sqrt{n}\cdot\nu\cdot10\log(d/d_{c})}{\sigma_{L}}\right),\label{eq:pi_fa_av_normal}
\end{equation}
for $d>d_{c}.$ This equation is a consequence of the scaling of $\eta_{L}$
and $\sigma_{L}^{2}$ by $n.$ Using the same distribution, but with
different parameters for $d<d_{c}$ is expected to be a worse match
to reality but allows to also evaluate the probability of missed detection
in closed from:
\begin{align}
\pi_{md,L,n}(d) & =\int_{-\infty}^{\langle\eta\rangle+\nu\cdot10\log(d/d_{c})}d\eta\,p_{L}(\eta)\nonumber \\
 & =1-Q\left(\frac{\sqrt{n}\cdot\nu\cdot10\log(d/d_{c})}{\sigma_{L}}\right)\nonumber \\
 & =Q\left(\frac{\sqrt{n}\cdot\nu\cdot10\log(d_{c}/d)}{\sigma_{L}}\right)\nonumber \\
 & =\pi_{fa,n}\left(\frac{d_{c}^{2}}{d}\right).\label{eq:pi_md_av_normal}
\end{align}
It leads to an interesting symmetry between the probabilities of missed
detection and of false alert.

Note that both probabilities $\pi_{md}$ and $\pi_{fa}$ depend on
the parameters of the distribution, on the true distance $d,$ and
on the critical distance $d_{c},$ but that they do not depend on
the explicit threshold $\Theta$, see Equation \eqref{eq:RSSI-Theta}
and the associated explanations. The resulting functional dependence
can either be used in a simulation of roaming users or can simply
be averaged over the interior of a circle of radius $d_{c}$ for $\pi_{md}$
or over its complement or a relevant subset for $\pi_{fa}$. The closed
form of Equation \eqref{eq:p_fa} provides the immediate insight that
$\pi_{fa,n}(d_{c})=1/2,$ which shows that the models are consistent
with our intuition.

\subsection{BLE Measurements Results\label{subsec:BLE-measurements}}

The companion paper by Dammann et al. \cite{DamGueGenGue20} describes
the measurements and their analysis in more details. All these measurements
have so far been made using ideal conditions with no additional people
except A and B (in the very initial measurements A was a actually
a post carrying the receiver). The experimental basis shall be further
broadened in the future. A first result can be derived from the estimated
Rice parameters at a distance of 2 meters $\gamma_{R}=247$ pW, and
$\sigma_{R^{2}}=9.15$ pW, as well as for the lognormal distribution
at 2 and 4 meters: 1.60 and $1.97$ dBm, respectively.

This allows plotting the functions from Equation \eqref{eq:pmd_d}
and \eqref{eq:pfa_d} for $\pi_{md,R,n}(d)$ and for $\pi_{fa,n}(d_{c}^{2}/d)=\pi_{md,L,n}(d)$,
respectively. The values of $n$ determines how many measurements
are combined into an elementary decision $\hat{c}_{1}$. For $n=1$,
the values $\pi_{md,R,1}(d)$ and $\pi_{fa,1}(d)$ are the best models
among those considered - the use of a decision threshold in the absolute
or logarithmic domain are equivalent. The parameter for 4 meters $1.97$
dBm is used for determining the false alarm rate.

\begin{figure}[tbh]
\begin{centering}
\includegraphics[width=8.5cm]{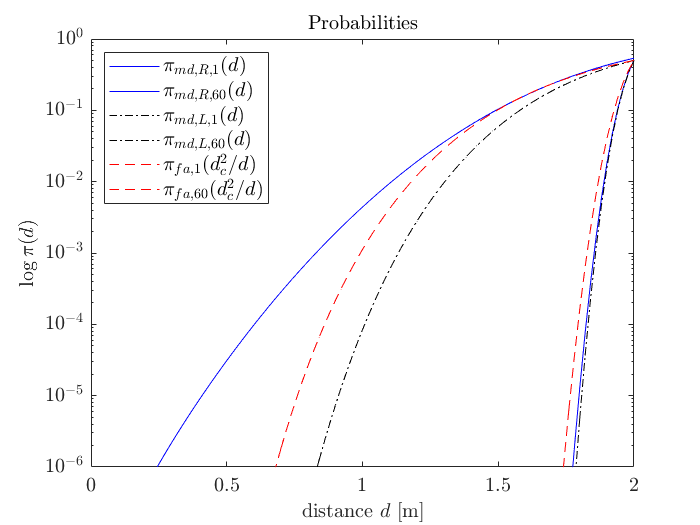}
\par\end{centering}
\caption{\label{fig:Timing-Infection}Probability of missed detection as a
function of user distance using Bluetooth Radio Signal Strength Indication
(RSSI).}
\end{figure}

If several RSSI values are added (logarithmic domain), the statistics
associated with the more realistic Rice distribution in the near range
can not be determined in closed form, at least not today. In this
case, Equation \eqref{eq:pi_md_av_normal} for the lognormal distribution
is used to determine $\pi_{md,L,n}(d)$ with the parameter for 2 meters.
This is used as an approximation of the true distribution in the exemplary
case $n=60$. The plots in Figure \eqref{fig:Timing-Infection} show
two groups of curves. The upper group corresponds to $n=1$ and the
lower group to $n=60.$ The latter group of curves shows the benefit
of diversity. Within these groups there are differences between $\pi_{md,R,n}(d)$
(wrong combination) and $\pi_{md,L,n}(d)$ (wrong fading statistics)
but they turn out not to be fundamental.

\begin{table}[tbh]
\caption{Selected values of $\pi_{md,av,n}$ under the assumption of a lognormal
fading distribution. These values define the number of measurements
needed to achieve the desired probability of further spreading the
disease.\label{tab:pi_md_av_n}}

\centering{}%
\begin{tabular}{|c|c||c|c|}
\hline
$n$ & $\pi_{md,av,n}$ & $n$ & $\pi_{md,av,n}$\tabularnewline
\hline
\hline
1 & 0.12 & 60 & 0.014\tabularnewline
\hline
6 & 0.054 & 120 & 0.007\tabularnewline
\hline
15 & 0.034 & 240 & 0.002\tabularnewline
\hline
30 & 0.023 & 480 & 0.0003\tabularnewline
\hline
\end{tabular}
\end{table}
In Section \ref{subsec:BLE-Propagation} the probability of missed
detection was determined as a function of distance. Since the probability
of detection is additive in the sense that:
\begin{align}
\pi_{d} & =\int dS(r)\,\rho(r)\,\pi_{d}(r).\nonumber \\
 & =\int_{0}^{d_{c}}2\pi rdr\,\rho(r)\,\pi_{d}(r)\label{eq:pi_d:marginalization}
\end{align}
In this equation $\pi_{d}(r)=1-\pi_{md}(r)$ is the condition probability
of detection given that fellow B is at distance $r$ and $dS(r)\,\rho(r)$
is the probability density for fellow B to be at that distance. Equation
\ref{eq:pi_d:marginalization} thus is the marginalization of $\pi_{d}(r)$
with respect to $r.$ Note that the limitation of the integration
is a consequence of $\pi_{d}(r)=0,$ whenever $r>d_{c}.$ This allows
to define the average probability of missed detection over the distribution
of users:
\begin{equation}
\pi_{md,av,n}=\frac{\int_{0}^{d_{c}}2\pi rdr\,\rho(r)\,\pi_{md,n}(r)}{\int_{0}^{d_{c}}2\pi rdr\,\rho(r)}.\label{eq:pi_md_av}
\end{equation}
The probability distribution of users in Equation \eqref{eq:pi_d:marginalization}
and \eqref{eq:pi_md} is given by:
\[
\rho(r)=\frac{dn(r)}{dS(r)}=\frac{d\lfloor\pi r^{2}\rfloor}{2\pi rdr}.
\]
In this expression $n(r)=\lfloor\pi r^{2}\rfloor$ is the number of
people at a distance not greater than $r$ in the case of a density
of one person per square meter. This corresponds to the densest packing
of people occupying a surface of 1 meter. People are continuously
spread in a symmetric manner around fellow A, which is a simple way
of achieving a densest packing. The ``function'' $dn(r)/dr$ is
mostly zero. It jumps at the values $r_{m}=\sqrt{m/\pi}$ with
\[
n(r_{m}+\epsilon)-n(r_{m}-\epsilon)=\int_{r_{m}-\epsilon}^{r_{m}+\epsilon}\frac{dn(r)}{dr}dr=1,
\]
which is a distribution in the sense of Schwartz \cite{Schw66}. With
these preparations, the integrals become:
\begin{equation}
\pi_{md,av,n}=\frac{1}{m_{c}}\sum_{m=1}^{m_{c}}\pi_{md,n}\left(\sqrt{\frac{m}{\pi}}\right),\label{eq:pmd_av}
\end{equation}
with $m_{c}$ being the largest integer with such that $r_{m_{c}}\leq d_{c}.$
Note that the density of points $r_{m}$ increases with increasing
$m,$ which means that the main contribution comes from the border
of the contact zone. Using the experimental results from \cite{DamGueGenGue20},
this integral is evaluated to $\pi_{md,av,1}=0.15$ for $n=1$ for
the $\chi^{2}$-distribution and to $\pi_{md,av,1}=0.12$ for the
lognormal distribution, which are both not very compatible with the
need of a small $x_{0}\pi_{md,}$. Remember that the latter value
is the reduction factor in the probability of further spreading of
the disease, achieved by contact tracing. Table \ref{tab:pi_md_av_n}
lists values of $\pi_{md,av,n}$, for different $n$, which can be
used to determine the reduction factor. Even in the case $n=120,$
the factor $x_{0}\pi_{md}=0.21$ in Model A and it would require 4
measurement per second. It is only with $n=480,$ that factor $x_{0}\pi_{md}$
falls below 1\%, which would require 16 measurements per second. This
would seriously impact the standby time of the smart phone. Assuming
Model C and a decision based on 3 minutes intervals, i.e. $x_{0}=5,$
means that we could achieve a reduction by a factor 0.07 provided
that $n=60$ measurements are performed and aggregated in each 3 minutes
interval, i.e. that one measurement is performed every 3 seconds.
In the case of a decision every 5 minutes, which assumes a lower dynamics
in the relative movement of people, the reduction factor is 0.04 with
the same 60 measurements, but now spread over a 5 minutes interval,
which corresponds to one measurement every 5 seconds. So, lower requirements
in the dynamic allow both to improve the suppression of the spreading
of the virus and to reduce the measurement rate.

Tolerable alarm rates were derived for the train scenario. This led
to the values in Table \eqref{tab:Indication for pfa}. The evaluation
of $\pi_{fa,n}(d)$ is straight forward. For $d=d_{c}$ it gives $\pi_{fa,n}(d_{c})=1/2$
as was already discussed previously. Assuming that people occupy a
circular surface of 1 square meter gives them a radius $\delta=1/\sqrt{\pi}.$
Thus, the minimum distance to people fully outside of the critical
zone is $d_{c}+\delta.$ Evaluating Equation \eqref{eq:pi_md_av_normal}
yields:
\[
p_{fa,1}(d_{c}+\delta)=0.137\quad\text{and\ensuremath{\quad p_{fa,3}(d_{c}+\delta)=0.029}}
\]
respectively. This means that $n=1$ is compatible with a journey
of 15 minutes before sending more than the two people to quarantine.
For $n=3$, long journeys of up to 3 hours become possible with the
same consequences. The probability of false alarm does thus not strongly
limit the number $n$ of measurements aggregated to a decision and
one might consider the more demanding homogeneous distribution of
users. This requires a study of the combination of false alarms. Consider
two fellows B and B', there is no alarm if neither B nor B' triggers
an alarm, i.e.:
\[
1-\pi_{fa}=(1-\pi_{fa,B})(1-\pi_{fa,B'}).
\]
Furthermore, let users be at distances $d_{c}+\delta(k+1)$ with $k\in\mathbb{Z^{+}}$
being a positive integer and assume that there are
\[
\nu(k)=\pi\left(d_{c}+2\delta(k+1)\right)^{2}-\pi\left(d_{c}+2\delta k\right)^{2}
\]
users at that distance (they cover an angular shell of thickness $2\delta$).
This guarantees a densest packing. In that case, the probability of
false alarm, i.e. an erroneous decision in favor of $C_{1},$ becomes:
\begin{equation}
p_{fa,n}=1-\prod_{k=0}^{\infty}\left(1-\pi_{fa,n}(d_{c}+k)\right)^{\nu(k)}.\label{eq:pfatot}
\end{equation}
In this more demanding scenario, exemplary values are:
\[
p_{fa,3}=0.413\quad\text{and\ensuremath{\quad p_{fa,9}=0.009,}}
\]
which means that $n=9$ would be sufficient to reduce the probability
of false alarm to a very small level. Table \ref{tab:Key-performance-parameters}
shows performance figures for a number of possible choices for the
number $n$ of measurements aggregated to an estimate $\hat{c}_{1},$
as well as for the number $x_{0}$ of estimates $\hat{c}_{1}$ that
lead to a decision $\hat{C}_{1}.$ The product of $n$ and $x_{0}$
leads to the measurement rate $\rho=x_{0}n/(15\cdot60).$ The performance
figures are the reduction factor $x_{0}\pi_{md,n}$ of the spreading
achieved by tracing as well as the probability of unduly sending a
person to quarantine. The figures in Table \ref{tab:Key-performance-parameters}
all relate to Model C. Model A does not lead to interesting parameter
choices and Model B is too static.

\begin{table}[tbh]

\caption{Key performance parameters: $x_{0}\pi_{md,n}$ measures the reduction
in spreading, and $p_{fa,n}$ the probability of undue quarantining.
The parameter $\rho$ is the number of measurements per second.\label{tab:Key-performance-parameters}}

\centering{}%
\begin{tabular}{|c|c|c|c|c|c|c|}
\cline{3-7} \cline{4-7} \cline{5-7} \cline{6-7} \cline{7-7}
\multicolumn{1}{c}{} &  & \multicolumn{2}{c|}{$x_{0}\pi_{md,n}$} & $p_{fa,n}$ & \multicolumn{2}{c|}{$\rho$}\tabularnewline
\cline{2-7} \cline{3-7} \cline{4-7} \cline{5-7} \cline{6-7} \cline{7-7}
\multicolumn{1}{c|}{} & $x_{0}$ & 3 & 5 & - & 3 & 5\tabularnewline
\hline
\multirow{3}{*}{$n$} & 6 & 0.16 & 0.27 & 0.064 & 1/50 & 1/30\tabularnewline
\cline{2-7} \cline{3-7} \cline{4-7} \cline{5-7} \cline{6-7} \cline{7-7}
 & 15 & 0.12 & 0.17 & 0.0002 & 1/20 & 1/12\tabularnewline
\cline{2-7} \cline{3-7} \cline{4-7} \cline{5-7} \cline{6-7} \cline{7-7}
 & 60 & 0.04 & 0.07 & 0.0000 & 1/5 & 1/3\tabularnewline
\hline
\end{tabular}
\end{table}

A choice with $n=15$ and $x_{0}=3$, for example, requires a measurement
to be performed every 12 seconds, suppressed the risk of spreading
by a factor 0.12 and does hardly send anyone unduly to quarantine.
Performing a measurement every five seconds reduces the risk of spreading
by a factor 0.04. This assumes that people let their phones hang from
their neck, and some standard form of environment. In reality, a number
of additional factors have to be taken into account, such as a more
complex propagation situation, e.g., due to metallic walls, a higher
dynamic of user movements, e.g. due to people entering and exiting
commuter trains, or unpredictable shadowing due to the user's hands,
arms or body in the path of radio signals. Thus, it is advisable to
complement the Bluetooth measurement by an alternative. Audio ranging
is the option that shall be described in the next section. The idea
is to use it whenever the situation is not clear.

\section{Audio Ranging\label{sec:Acoustic-Ranging}}

\begin{figure}[tbh]
\centering{}\includegraphics[width=8.5cm]{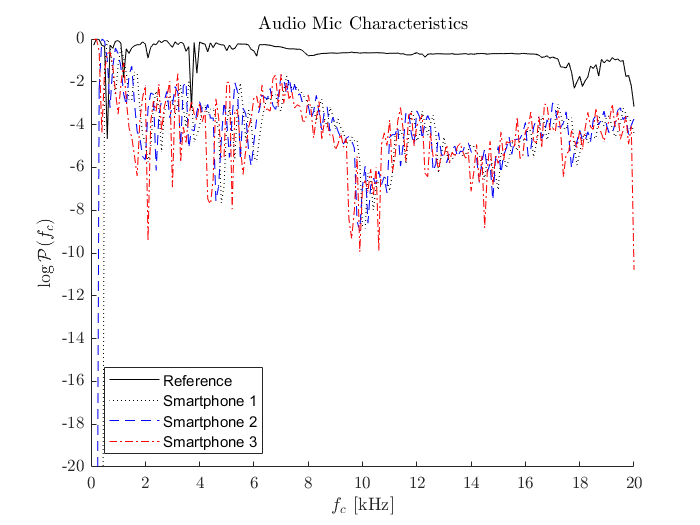}\caption{\label{fig:Microphones}Frequency response of microphones from three
different smartphones.}
\end{figure}

Smartphones have a microphone and a speaker with rather good transmit
and receive conditions if the device is carried on the chest or held
in the hand. This can be used for audio ranging up to distances of
a few meters. Signals and their transmission can be configured by
the API. In experiments that we performed recently, we focused on
the use Android phones. The response of the microphones built into
three different phones is shown in Figure \ref{fig:Microphones}.
The references were a NT1-A microphone from Rode and an Adagio Infinite
Speaker of A3 on the source side. Figure \ref{fig:Microphones} shows
the response of three smartphones from two different brands. The curves
are very similar, suggesting that the same microphones are integrated
in those phones. All microphones show a good sensitivity over all
frequencies.

A similar experiment was performed for the speakers with a rather
different result. In that case only two smartphones were analyzed.
The response on the better device is reduced by roughly 10 dB above
16 kHz, as compared to the reference. The response of the other one
is degraded by another 3 dB and the degradation starts 2kHz earlier.
Covering the speaker by one layer of tissue of a sweater degrades
the performance by another 4 dB. If both parties cover their smartphones
the associated attenuation adds up. Thus, the use of audio ranging
requires carrying the devices in an exposed manner, e.g. hanging from
one's neck, see Figure \eqref{fig:Handykette}. Transmission at lower,
less attenuated frequencies is not considered as a true option, since
it would be too disturbing. The norm ISO 226:2003 compiles equivalent
hearing sensitivity (isophones), which allows to compare the disturbance
caused by acoustical signals on different frequencies.

\begin{figure}[tbh]
\begin{centering}
\includegraphics[width=8.5cm]{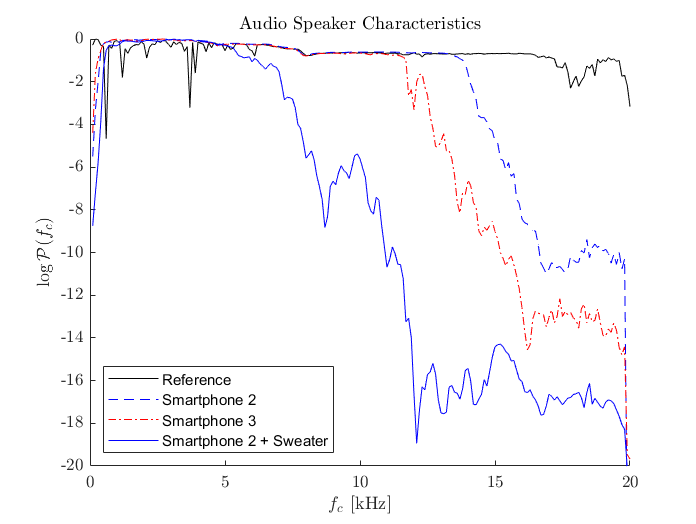}
\par\end{centering}
\caption{\label{fig:Speakers}Frequency response of two speakers, as well as
influence of covering the speaker of Smartphone 2 with one layer of
a sweater.}
\end{figure}

On the basis of such considerations, we propose modulating a carrier
at 18 kHz with a modulation rate of 1 kbaud. This keeps the signal
in a spectral range that is not too disturbing to most people. A spread
spectrum modulation provides a good range resolution and allows to
operate at a low signal-to-noise ratio at the same time. Different
options exist and are discussed in \cite{KurGueGue20}. Since the
velocity of sound in air is $c_{s}=343$ m/s under standard conditions,
a chip duration of 1 ms corresponds to a length of 34 cm. At a typical
signal-to-noise ratio this leads to a distance resolution of 1 to
3 cm. Let us be conservative and assume a resolution of 5 cm. A multipath
delay of two meters leads to an offset by 6 chips and is well suppressed
by the autocorrelation of the spreading code. The length of the spreading
code is assumed to be around 350 chips. An alternative using chirps
is also considered. The performance of audio ranging is further developed
in Section \ref{sec:Acoustic-Ranging}.

Audio ranging can be performed in a peer-to-peer or in a networked
manner. Consider the peer-to-peer situation first. Smartphones do
not provide accurate timing control. However, the microphone input
of a smartphone may be sampled at a fixed rate. Furthermore, smartphones
can transmit and receive at the same time and this is furthermore
supported by the APIs of Android and iOS. Let the smartphones thus
agree to start audio ranging via Bluetooth . In a first step they
open their microphone channels and then proceed according to Figure
\ref{fig:Acoustical-ranging}: at time $t_{TX,A}$, A transmits the
ranging signal using its speaker. This transmission is delayed with
respect to the API by $\tau_{TX,A}$. In parallel to its transmission
A's microphone capture the transmitted signal. This signal is delayed
by the sum of the local propagation delay $\tau_{l,A}$ and by the
internal receive delay $\tau_{RX,A}.$ The delay $\tau_{l,A}$ is
determined by the device geometry and can be stored in memory. A standard
value of 14 cm should be appropriate for most devices on the market.
The time of reception thus is:
\[
t'_{RX,A}=t_{TX,A}+\tau_{TX,A}+\tau_{l,A}+\tau_{RX,A},
\]
and is used for calibration purposes. The same definition of delays
applies at B. Thus, the signal transmitted by A at time $t_{TX,A}$
is received at B at the time $t_{RX,B}$:

\[
t_{RX,B}=t_{TX,A}+\tau_{TX,A}+\tau+\tau_{RX,B},
\]
with $\tau$ being the propagation time from A to B. After reception
of the signal from A by B, B sends a corresponding signal to A. The
equations are obtained by changing the roles of A and B:

\[
t'_{RX,B}=t_{TX,B}+\tau_{TX,B}+\tau_{l,B}+\tau_{RX,B},
\]
and
\[
t_{RX,A}=t_{TX,B}+\tau_{TX,B}+\tau+\tau_{RX,A}.
\]
At the end of the reception A sends
\begin{equation}
\Delta t_{A}=t_{RX,A}-t'_{RX,A}+\tau_{l,A}\label{eq:Delta_time}
\end{equation}
to B and B sends $\Delta t_{B}=t_{RX,B}-t'_{RX,B}+\tau_{l,B},$ using
BLE. Thus, both can compute the propagation time:
\[
\tau=\frac{\Delta t_{A}+\Delta t_{B}}{2},
\]
and thus the distance $d=\tau c_{s}.$ The property of audio signals,
which is crucial for this self-calibration, is the possibility to
observe the own transmitted signal.
\begin{center}
\begin{figure}[tbh]
\centering{}\includegraphics[width=8.5cm]{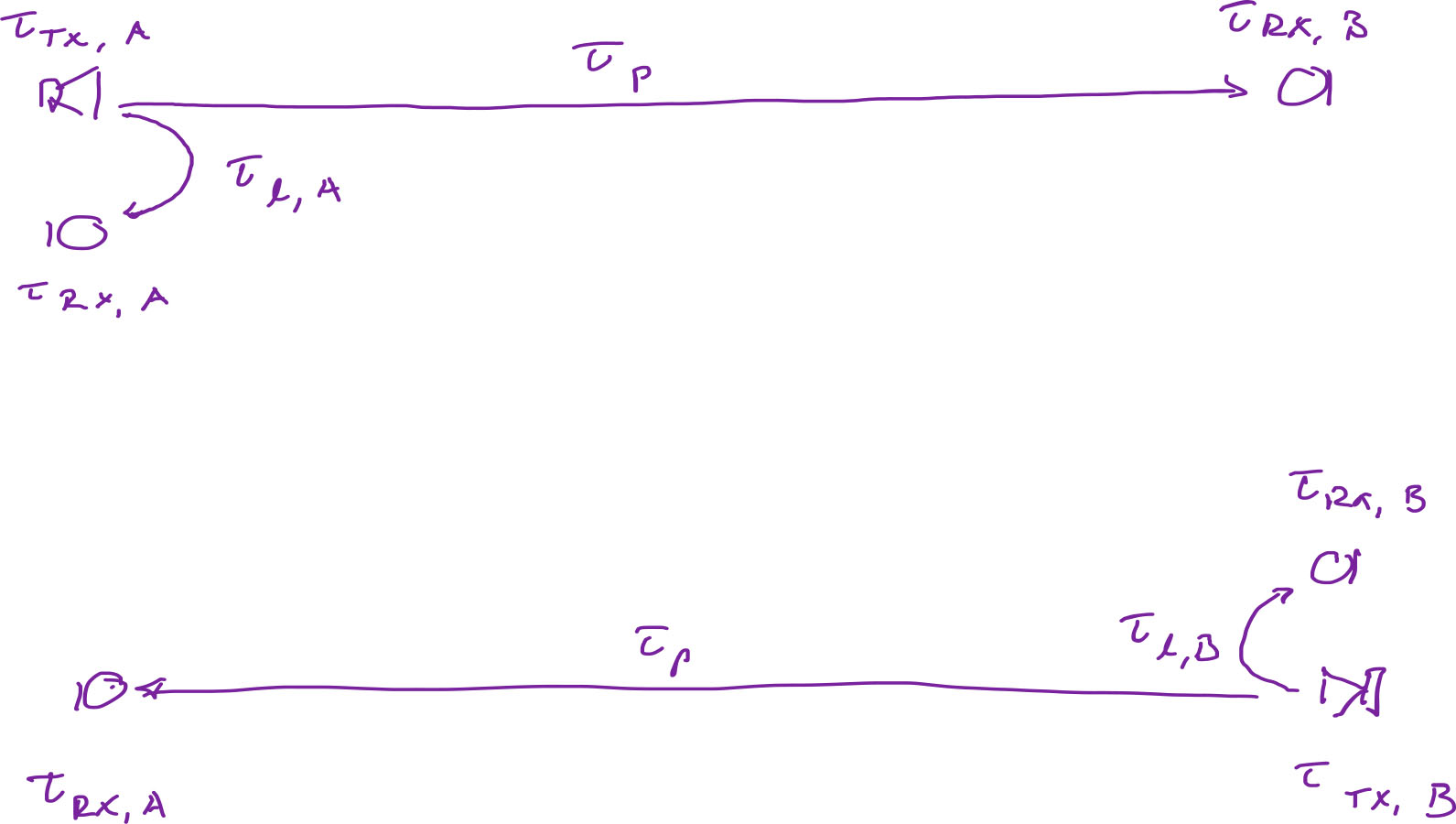}\caption{Signal paths in two-way acoustical ranging with calibration of transmit
and receive delays.\label{fig:Acoustical-ranging}}
\end{figure}
\par\end{center}

\subsection{Ranging Protocol\label{subsec:Ranging-Protocol}}

The above peer-to-peer protocol can be extended to a networked protocol.
In this case, the users agree on an ordering of transmissions via
Bluetooth. All smartphones $A_{1}\ldots A_{k}$ activate their microphones
and one after the other transmit their audio ranging signals. For
simplicity, the scheduling is prearranged, which also works if some
of the smartphone cannot acquire all signals. In this case, all delays
are summed up: 350 ms for the ranging signal, 10 ms (corresponding
to 4 meters) for propagation and 40 ms for the internal delays between
the activation of the transmission command and the start of transmission
(the latter needs to be confirmed by more data). This allows for a
scheduling of a transmission every 400 ms. After the completion of
the cycle and the evaluation of the reception time $t_{RX,A_{i}}$
by terminal $A_{1}$, this terminal transmits the time difference
using Bluetooth:
\[
\Delta t_{A_{1},A_{i}}=t_{RX,A_{i}}-t'_{RX,A_{1}}+\tau_{l,A_{1}}\quad\text{for \ensuremath{2\leq i\leq k,}}
\]
If all terminals see each other, they transmit $k(k-1)$ such values
in total. The annoying transmissions of audio signals remain limited
to $k$, however. The overall time interval spanned by all transmissions
in the networked protocol may be long enough for users to move slightly.
This is not critical, however. The snap-shot measurements are simply
converted to average values. The only instances, which require some
care are those in which the audio signals are used to calibrate Bluetooth
measurements. Finally, it should be emphasized that audio beacon transmissions
should not be activated if the device is held to the ear. Even if
the signals are hardly heard, this seems a reasonable precaution.

\subsection{Theoretical Performance of Acoustic Range Estimation\label{subsec:Acoustic-Performance}}

\begin{figure}[tbh]
\begin{centering}
\includegraphics[width=8.5cm]{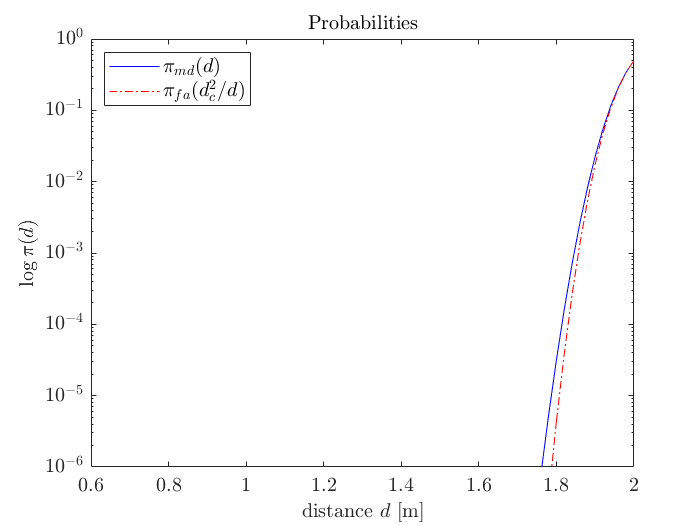}
\par\end{centering}
\caption{Probabilities of missed detection as a function of $d$ and of false
alarm as a function of $d_{c}^{2}/d$ for audio ranging.\label{fig:Probabilities-audio-ranging}}

\end{figure}

The received audio signal is filtered to remove out-of-band interference
and noise to the best possible extent. The filtered signal is used
to determine the in-band interference and noise level ${\cal N}_{0}$
and is furthermore correlated using the filtered ranging signal. For
simplicity, the further exposition focuses on spread spectrum signals.
In a first step the I and Q components of the correlation $C(\Delta\tau)$
are computed at intervals of $T_{c}/2$ with $T_{c}=1$ ms denoting
the chip duration. The result is searched for the delay leading to
the maximum norm $|C(\Delta\tau)|$. Although, the implementations
by widely used phones seem not to require that, frequency offsets
may be searched as well. This allows to acquire the signal which may
be present or not. Thus, it is sufficient to search for the delay
(and frequency offsets) leading to the maximum norm from early to
late. The latter ordering is to avoid locking on an echo. If the signal
to noise ratio is above the expected threshold, the signal is assumed
present. In this case, a successive refinement of the result is performed
in a DLL type of processing. The power discriminator
\[
D_{P}(\Delta\tau)=|R(\Delta\tau+\delta)|^{2}-|R(\Delta\tau-\delta)|^{2}
\]
is used to iteratively increase/reduce the delay $\Delta\tau$ depending
on the value of $D_{P}(\Delta\tau)\gtrless0.$ In this equation $\delta$
is half the correlator spacing and is expressed as a fraction $\Delta$
of the chip duration: $\delta=\Delta T_{c}$. We will restrict ourselves
to $\Delta=1.$ A further optimization is possible, see Betz and Kolodziejski
\cite{BetKol09,BetKol09a}. The uncertainty of the delay estimate
$\Delta\tau$ due to noise is given by (see Dierendonck, Fenton and
Ford \cite{DieFenFor92}):
\begin{equation}
\sigma_{\Delta\tau}^{2}\simeq T_{c}^{2}\frac{\Delta}{4{\cal E}_{i}/{\cal N}_{0}}\left(1+\frac{3}{(2-\Delta){\cal E}_{i}/{\cal N}_{0}}\right).\label{eq:Audio_Uncertainty}
\end{equation}
In this expression, ${\cal E}_{i}$ is the signal energy accumulated
during the correlation, and ${\cal N}_{0}$ is the spectral noise
density of the audio noise and interference. The latter quantity is
estimated using the norm of the filtered I and Q samples of the incoming
signal:
\[
{\cal N}_{0}=\frac{1}{B_{S}NT_{c}}\sum_{n=1}^{N}\left(s_{I}^{2}+s_{Q}^{2}\right),
\]
with $N$ denoting the number of samples and with $B_{S}$ denoting
the bandwidth of the passband filter. This estimate is performed ahead
of time and is used for setting the volume of the transmission, such
that ${\cal E}_{i}/{\cal N}_{0}$= 6 dB at 4 meters. At this level
the signal can be acquired, and Equation \eqref{eq:Audio_Uncertainty}
implies that $\sigma_{\Delta\tau}\simeq T_{c}/4,$which corresponds
to 9 cm. At 2 meters, this is half that value, i.e. 4.5 cm. The calibration
of the transmit power may be performed by listening to the own beacon.
This allows detecting whether the user is inadvertently covering the
microphone or the speaker, which should trigger a request to the user
to remove the blockage. The distribution of audio ranging measurements
is Gaussian with a standard deviation given by Equation \eqref{eq:Audio_Uncertainty}.
This allows computing $\pi_{md}$, i.e the probability of deciding
against $\hat{c}_{1}$, as a function of the distance $d\leq d_{c}$:
\begin{align}
\pi_{md}(d) & =\int_{d_{c}}^{\infty}dx\frac{1}{\sqrt{2\pi}\sigma_{\Delta\tau}}e^{-(x-d)^{2}/(2\sigma_{\Delta\tau})}\nonumber \\
 & =Q\left(\frac{d_{c}-d}{\sigma_{\Delta\tau}}\right),\label{eq:pmd_audio}
\end{align}
and $\pi_{fa}$, i.e the probability of wrongly deciding in favor
of $\hat{c}_{1}$, for distances $d>d_{c}$:
\begin{align}
\pi_{fa}(d) & =\int_{0}^{d_{c}}dx\frac{1}{\sqrt{2\pi}\sigma_{\Delta\tau}}e^{-(x-d)^{2}/(2\sigma_{\Delta\tau})}\nonumber \\
 & =1-Q\left(\frac{d_{c}-d}{\sigma_{\Delta\tau}}\right).\label{eq:pfa_audio}
\end{align}
Note that the symmetry of lognormal fading between $\pi_{md}(d)$
and $\pi_{fa}(d_{c}^{2}/d)$ is lost. The plot for audio ranging,
corresponding to $\sigma_{\Delta\tau}=5$ cm is shown in Figure \ref{fig:Probabilities-audio-ranging}

Again one might evaluate the average rate of missed detection and
of false alarm as in Equation \eqref{eq:pmd_av}. In this case, the
averaged probability of missed detection becomes $\pi_{md,av}=0.016$.
In the present case, the number of measurements is primarily limited
by the acoustical disturbances associated with the transmission of
the beacon. The number of measurements $n$ used for taking a decision
is always 1. Furthermore, the number of measurements $x_{0}$ per
15 minutes must also be small for the same reason. With $x_{0}=3,$
the reduction of the spreading rate of disease is $x_{0}\pi_{md,av}<0.05,$
which is a low figure. The probability of false alarm described by
Equation \eqref{eq:pfa_audio} decays so quickly that it is insignificant
at $d=d_{c}+\delta,$ i.e. $\pi_{fa}(d_{c}+\delta)\simeq0.$ The same
applies for the integration over a two-dimensional plane according
to Equation \eqref{eq:pfatot}.

The present discussion was about the contributions of uncertainty
due to signaling. Additionally, the relative geometry of the microphones
and speakers may add some bias, which may lead to a shift of the border
to a contact zone by a few centimeters. This is rather uncritical,
however. The important conclusion is that audio ranging provides sharp
results. This form of ranging might thus be activated whenever the
information gained by Bluetooth measurements may lead to a wrong conclusion.

\section{Attitude Sensing\label{sec:Attitude-Sensing}}

This section is more a reference to options that may be considered.
The benefits will become visible by the qualitative discussion of
Section \ref{sec:Classification}. Earth gravity in the $-\vec{e_{z}}$
direction, i.e. towards the center of the earth and the magnetic field
in the direction of $\vec{e}_{mN},$ i.e. towards magnetic North provide
two directions that enable attitude determination. Both are seriously
disturbed in ways that depend on the environment. A number of authors
have investigated the quality of attitude sensing both using algorithms
built into smartphones and using own estimation algorithms. Michel
and co-authors summarize a number of findings \cite{MicGenFouLay17}.
They report an accuracy of $6^{\circ}$ with a sampling rate of 40Hz
whenever the smartphone is kept in a relatively calm position (front
pocket, texting or phoning). These results apply to their own algorithms
``MichelObsF'' and ``MichelEkfF.'' They did not study the behavior
in a train, which is a particularity difficult environment: with many
sources of acceleration, due to the track geometry, due to passing
switches or simply due to irregularities in the tracks themselves.
Similarly, the magnetic field in trains is modulated by electrical
motors, permanent magnets and large currents. On the other hand people
sitting or standing next to each others are likely to be affected
in a similar manner. Exploiting the latter property, however, requires
the use of common standardized algorithm and precise time stamping
of measurements.

Carrying the smartphone by letting it hang down one's neck leads to
two stable orientation, one with the display facing the chest and
one with the display facing ahead. The resolution of the associated
ambiguity is rather straight-forward, at least as long as people do
not predominantly walk backward. Alternatively, the cameras could
be used for determining the orientation, since the brightness of the
pictures is very different. Pitch angles are suppressed by gravity,
as long as people do not bend backwards, which is unnatural. Roll
angles may occur if one strap is shorter than the other one. They
are compensated by sensing earth gravity. In our opinion the context
of COVID-tracing is quite favorable to the use of relative attitude
estimation, which would provide an interesting complement to Bluetooth
sensing and/or acoustic ranging. This needs to be developed, however.

\section{Classification\label{sec:Classification}}

The definition of a Category 1 contact by the Robert Koch Institute
\cite{Category20} includes three elements:
\begin{itemize}
\item an accumulated duration of 15 minutes, which can easily be metered,
\item a distance of less than 2 meters, which is more difficult to establish,
\item and the concept of being face-to-face, discussed below.
\end{itemize}
From the previous sections, specially Section \ref{sec:Statistical-Considerations}
and \ref{sec:Power-Measurements}, we learned that under idealized
conditions, Bluetooth RSSI measurements provide an adequate estimation
of the distance between two fellows or more exactly an estimate on
whether B is in the critical zone of A. The probability of missed
detection was found to the be a critical performance measure. Audio
ranging was found to be an interesting complement to Bluetooth measurements,
in particular if the latter measurements are disturbed by shadowing
or multipath. They provide a comparatively sharp answer, and may be
used to calibrate past and future Bluetooth RSSI measurement. Audio
measurements may be audible and thus annoying for younger people,
as well as for dogs and other animals. As a consequence, it is beneficial
to keep them sparse. In Section \ref{sec:Attitude-Sensing}, we very
shortly addressed the use of attitude sensing.

\begin{figure}[h]
\begin{centering}
\includegraphics[viewport=0bp 500bp 1800bp 2500bp,clip,width=8.5cm]{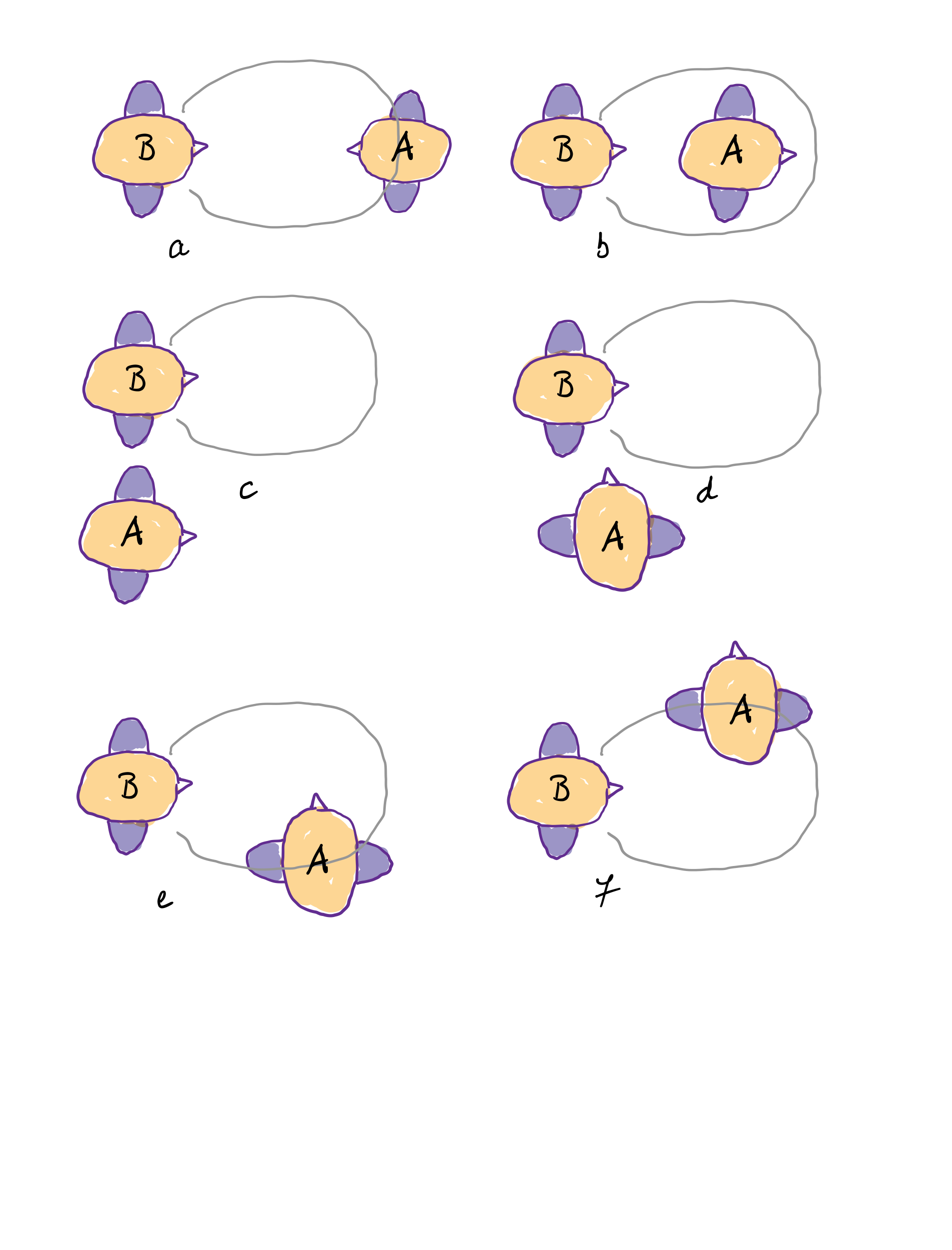}
\par\end{centering}
\caption{\label{fig:Different-poses}Different potential poses of a COVID-19
carrier A and of a nearby person B. The bubble in front of A shows
the area into which A exhales air carrying droplets with the virus.}
\end{figure}

In this section, we shall superficially address the potential of combining
these measurement types. For this discussion, it is meaningful to
differentiate different poses, as shown in Figure \ref{fig:Different-poses}.
A selection of essential poses of two fellows in close proximity is
shown in a top view. Fellow B is infected and exhales air charged
with microscopic droplets carrying the virus. Fellow A inhales the
droplets. Pose (a) in Figure \ref{fig:Different-poses} is what everyone
would agree to call a face-to-face situation. It is the type of situation,
which occurs during a meeting, lunch or in public transportation for
people sitting or standing opposite to each other. It might also occur
when desks are facing each other and in some other special situations.
Pose (b) occurs in public transportation, in queues as well as in
lecture halls, concert halls, cinemas or the like. It also appears
dangerous, although Fellow B needs to be closer for that, but this
might often be the case. However, unless B stands and is much taller
than A, the air flow will only partially reach A's nose and mouth.
A further specification by medical authorities would be helpful in
this case. Pose (c) occurs in similar situations as Pose (b). Pose
(d), (e) and (f) occur during meetings both while standing and sitting,
in public transportation and some other situations. Pose (c) and (d)
do not appear too critical, although B is likely to turn his head
from time to time, which is not detected by the sensors considered.
Pose (d), (e), and (f) are difficult to differentiate even using perfect
ranging and orientation.

Assuming that there is no specific direction in the air-flow, due
to wind or draft, and that the different poses can be differentiated,
medical requirements would probably choose
\begin{itemize}
\item Pose (a), (d), and (e) to be Category 1, i.e. critical,
\item Pose (b) would be critical for a lower distance which might depend
on the height differences,
\item Pose (c) and (f) would be essentially uncritical.
\end{itemize}
The possibility to discriminate the cases depends on the type of sensing,
as described so far, and is discussed in the following three sections.

\subsection{Bluetooth-only Measurements}

BLE RSSI measurements will return similar results for the Poses (a),
(c), (d), and (e). The distance $d$ between the fellows might appear
larger in Pose (f) than it actually is. This is uncritical, however.
In Pose (b), the received power will be associated with a larger distance
than the actual one, as well. Depending on how Pose (b) is classified,
this leads to a missed detection. A similar situation may also occur
in Pose (e) whenever Fellow A obstructs the line of sight with his
left harm, e.g. by holding himself on a bar in public transportation.
All missed detection events are critical since they leave close encounters
undetected. Finally, the poses (c) and (d) will typically generate
false alarms, which sends people to quarantine and testing. This sort
of differentiation has not been considered so far, at least to our
knowledge.

\subsection{Bluetooth, Attitude Sensing}

The addition of a attitude sensing, allows to separate the cases of
``Pose (b) with a small distance'' from ``Pose (a) with a large
distance''. Thus, it might use a lower threshold in the case of an
aligned attitude and thus avoid the missed detection events in Pose
(b). With a lower threshold, however, fellows in Pose (c) will be
identified as $C_{1}$ up to a rather large relative distance, potentially
generating many false alarms.

\subsection{Bluetooth, Attitude Sensing and Audio Ranging}

An extensive use of audio ranging, would eliminate false alarms mostly.
It would implement the conditions of Category 1 without the alleviation
due to the the condition of being face-to-face. When combined with
the other measurements, audio measurements provide additional discrimination
and allow reducing the rate of missed detection and false alarms.
In reality, acoustical signals are subject to multipath, which might
be critical if the direct path is strongly attenuated. Since the receiver
searches from early to late it is unlikely to be induced in error,
however, as long as the direct path can still be detected.

\section{Conclusions}

Difficulties in Bluetooth RSSI-based ranging are mentioned by a number
of scientists orally. The significant attenuation by the human body
and other influencing factors, such as keys, coins, metallic pens,
business card holders and the like make the power levels very unpredictable.
We thus propose to standardize the wearing of smartphones or alternative
devices on the chest, when not held in the hand or used for making
phone calls. This provides an environment that is much better defined
for Bluetooth RSSI-based ranging, audio ranging and attitude determination.
Currently, we don't see an alternative setting to the present one
that allows for an analysis of the tracing performance in terms of
identifying Category 1 contacts and avoiding unduly frequent alerts
for contacts that are not Category 1. The analysis shows that the
accumulated statistics require low figures for the per event missed
detection rate. This can be achieved with measurements every few seconds
aggregated into decisions every few minutes, which is adequate for
stable distributions of people, such as in a meeting, at lunch, in
a train and the like. The false alarm rate is a lesser problem as
soon as a few measurements are aggregated. The analysis presented
in the paper is a preliminary one. Much more experimental data should
be generated to refine the findings. In Germany, the current probability
of encountering an infected person is rather low. In such a context
the performance does not matter too much. There are many regions in
the world, where this is not the case, however. It would thus be quite
beneficial if this work was taken up and further developed, in particular
with respect to attitude sensing. Some individuals may reject the
idea of carrying their smartphone around their neck. This could be
addressed by producing decorative gadgets which are less obstructive
to wear. Beyond that the carrying of a device around the neck also
enables the use of the camera. This would allow to further refine
the evaluation of the risk but would drain the batteries much more
and would raise concerns about privacy Thus, the use of the sensors
addressed in the present papers seem to remain most promising. In
the future, Bluetooth ranging should be considered as well. The complete
analysis of the paper and its validity rely on the current model of
infection of the Robert-Koch Institute.

\section*{Acknowledgment}

The authors would like to thank Dr. Armin Dammann from the German
Aerospace Center (DLR) for comments on Section \ref{subsec:BLE-Propagation}
and for providing us with early results from the evaluation of the
experiments.

\section{References}

\bibliographystyle{IEEEtran}
\bibliography{Bibliography}

\end{document}